\documentclass[twocolumn,epsfig,aps,prb]{revtex4}

\begin{document}

\title{Optical models of the big bang and non-trivial space-time metrics based on metamaterials}

\author{Igor I. Smolyaninov}
\affiliation{Department of Electrical and Computer Engineering, University of Maryland, College Park, MD 20742, USA}

\date{\today}

\begin{abstract}
Optics of metamaterials is shown to provide interesting table top models of many non-trivial space-time metrics. The range of possibilities is broader than the one allowed in classical general relativity. For example, extraordinary waves in indefinite metamaterials experience an effective metric, which is formally equivalent to the \lq\lq two times physics \rq\rq model in 2+2 dimensions. An optical analogue of a \lq\lq big bang\rq\rq event is presented during which a (2+1) Minkowski space-time is created together with large number of particles populating this space-time. Such metamaterial models enable experimental exploration of the metric phase transitions to and from the Minkowski space-time as a function of temperature and/or light frequency.
\end{abstract}


\maketitle

Unprecedented degree of control of the local dielectric permittivity 
$\epsilon _{ik}$ and magnetic permeability $\mu _{ik}$ tensors in novel electromagnetic metamaterials has fueled recent explosion in novel device ideas, and resulted in discovery of new physical phenomena. Advances in experimental design and theoretical understanding of electromagnetic metamaterials greatly benefited from the field theoretical ideas developed to describe physics in curvilinear space-times. Electromagnetic cloaking \cite{1,2,3} and electromagnetic wormholes \cite{4} may be cited as good examples. On the other hand, it should be noted that compared to standard general relativity, metamaterial optics gives an experimentalist much more freedom to design an effective space-time with very unusual properties. The flat Minkowski space-time with the usual (-,+,+,+) signature does not need to be a starting point. Other effective signatures, such as the \lq\lq two times physics \rq\rq \cite{5} (-,-,+,+) signature may be realized in the experiment, and the spatio-temporal symmetries and physical properties of such an unusual highly symmetric metamaterial space can be explored. To illustrate this point, let us consider a non-magnetic ($\mu =1$) uniaxial anisotropic metamaterial with dielectric permittivities $\epsilon _x=\epsilon _y=\epsilon _1$ and $\epsilon _z=\epsilon _2$. The wave equation for the extraordinary wave (see for example \cite{6}) is  

\begin{equation}
\label{eq1}
\frac{\partial ^2 E}{c^2\partial t^2}=\frac{1}{\epsilon _1}\frac{\partial ^2 E}{\partial z^2}+\frac{1}{\epsilon _2}(\frac{\partial ^2 E}{\partial x^2}+\frac{\partial ^2 E}{\partial y^2}) 
\end{equation}

where $E$ is the electric field amplitude of the extraordinary wave. In ordinary crystalline anisotropic materials both $\epsilon _1$ and $\epsilon _2$ are positive. On the other hand, in the so-called indefinite metamaterials considered for example in \cite{7,8} one could have $\epsilon _1<0$ while $\epsilon _2>0$, or the other way around. In the visible frequency range these metamaterials are typically composed of multilayer metal-dielectric or metal wire array structures \cite{8}. Optical properties of such metamaterials are quite unusual. For example, it was demonstrated theoretically in \cite{7,9,10} that there is no usual diffraction limit in an indefinite metamaterial. This prediction has been confirmed experimentally in \cite{11,12}. 

Let us consider the case of $\epsilon _1<0$ and $\epsilon _2>0$. Let us assume that this behavior holds in a wide enough frequency range, and that the metamaterial dispersion may be neglected in the same frequency range. As a result, we observe that the effective space-time signature as seen by extraordinary light propagating inside the metamaterial has different character at different space-time scales. At high frequencies (above the plasma frequency of the metal) which corresponds to the small space-time scale, the metamaterial exhibits \lq\lq normal \rq\rq Minkowski effective metric with 
(-,+,+,+) signature. On the other hand, at low frequencies (at large space-time scale) the metric signature changes to (-,-,+,+), and has the \lq\lq two times physics \rq\rq character. After simple coordinate transformation eq.(1) can be re-written as

\begin{equation}
\label{eq2}
(\frac{\partial ^2}{\partial x_0^2}+\frac{\partial ^2 }{\partial x_3^2})E=(\frac{\partial ^2}{\partial x_1^2}+\frac{\partial ^2}{\partial x_2^2})E , 
\end{equation}

which explicitly demonstrates highly symmetric nature of the system: the wave equation exhibits rotational symmetry both in the $(x_1,x_2)$ and in the $(x_0,x_3)$ planes. In addition, since the $(x_1,x_2)$ and $(x_0,x_3)$ pairs are interchangeable, for the extraordinary light in the low frequency and the small boost limit there is no formal difference between the original $x_0$ time coordinate and any of the spatial directions. 

Alternatively, in the case of $\epsilon _1>0$ and $\epsilon _2<0$, eq.(1) can be re-written as 

\begin{equation}
\label{eq3}
(\frac{\partial ^2}{\partial x_0^2}+\frac{\partial ^2 }{\partial x_1^2}+\frac{\partial ^2}{\partial x_2^2})E=(\frac{\partial ^2}{\partial x_3^2})E  
\end{equation}

As a result, both at the small and the large space-time scales the effective metric looks like the Minkowski space-time. However, at large scale the $x_3$ coordinate assumes the role of a time-like variable. Note that causality and the form of eqs.(2) and (3) place stringent limits on the material losses and dispersion of hyperbolic metamaterials: a dispersionless and lossless hyperbolic metamaterial would violate causality. On the other hand, such metamaterials enable experimental exploration of the \lq\lq metric phase transitions \rq\rq to and from the Minkowski space-time as a function of the light frequency (the space-time scale).

The described behavior is not limited to the artificial metal-dielectric metamaterials only. Many dielectric crystals, such as $\alpha$-quartz, have lattice vibration modes which carry an electric dipole moment. The dipole moment couples the lattice vibrations to the radiation field in the crystal to form the so-called phonon-polariton modes \cite{13}. As a result, multiple so-called reststrahlen bands are formed near the frequencies of the dipole-active lattice vibration modes $\omega_n$ (where $n$ is the mode number) in which both $\epsilon _1$ and $\epsilon _2$ become metal-like and negative. These bands are typically located in the mid-IR region of the electromagnetic spectrum. However, due to crystal anisotropy, $\epsilon _1$ and $\epsilon _2$ change sign at slightly different frequencies \cite{14}. Thus, in the narrow frequency ranges near the boundaries of the reststrahlen bands $\epsilon _1$ and $\epsilon _2$ have different signs. In these narrow frequency ranges natural crystals behave as indefinite metamaterials, and the extraordinary light dispersion law looks either like $k_x^2+k_y^2-k_z^2=\omega_n^2$ or $k_z^2-k_x^2-k_y^2=\omega_n^2$. The latter case corresponds formally to the free particle spectra in a (2+1) dimensional Minkowski space-time, in which the mass spectrum is given by the spectrum $\omega_n$ of the lattice vibrations. It may be considered as an experimental model of a \lq\lq two times physics\rq\rq system \cite{5} in which one time-like coordinate is compactified (however, due to causality issues this analogy remains formal). In principle, the lattice symmetry group and the lattice eigenmodes may be adjusted to reproduce any desired mass spectrum. 

It is interesting to note that the liquid-solid phase transition in such a crystal provides us with an example of a phase transition in which a formal (2+1) dimensional Minkowski space-time emerges together with a discrete free particle spectrum. The characteristic feature of this phase transition appears to be a kind of toy \lq\lq big bang\rq\rq due to the sudden emergence of the infinities of the photonic density of states near the $\omega_n$ frequencies. Unlike the usual black body photonic density of states 

\begin{equation}
\label{eq4}
\frac{dn}{d\omega }=\frac{\omega ^2}{2\pi ^2c^3} 
\end{equation}
   
defined by the \lq\lq normal\rq\rq $k_x^2+k_y^2+k_z^2=\omega ^2$ photon dispersion law, the density of states of the extraordinary photons near the $\omega_n$ frequencies diverges in the lossless continuous medium limit:

\begin{equation}
\label{eq5}
\frac{dn}{d\omega }=\frac{K_{max}\omega }{\pi ^2c^2} , 
\end{equation}
 
where $K_{max}$ is the momentum cutoff. This momentum cutoff must be introduced because in the lossless continuous medium approximation the absolute value of the $k$-vector projection on any axis can be arbitrary large. This is the physical reason for the absence of the diffraction limit in an indefinite metamaterial \cite{9,10,11,12}. As a result of the suddenly emerging divergencies in the photonic density of states near the $\omega_n$ frequencies, these states are quickly populated during the liquid-solid phase transition. This event can be considered as a simultaneous emergence of the (2+1) dimensional Minkowski space-time and a very large number of particles (extraordinary photons), which quickly populate the divergent density of states, and hence the emergent (2+1) Minkowski space-time itself. This consideration provides us with another example of an interesting analogy between the metamaterial optics and the gravitation theory. This analogy may prove to be as useful as the well-known analogy between the physics of the superfluid states in He$^3$ and the gravitation theory \cite{15}.

In addition to two possibilities mentioned above, we should emphasize that the ease of control of the dielectric permittivity $\epsilon _{ik}$ and magnetic permeability $\mu _{ik}$ tensors in metamaterial optics lets an experimentalist create optical models of virtually any metric allowed in general relativity. Recent paper by Genov $et$ $al.$ \cite{16} provides additional examples.

\end{document}